# PLANNING NEAR EARTH ASTEROID OBSERVATIONS ON A 1m CLASS TELESCOPE

OVIDIU VADUVESCU

*ACRU & SAAO, South Africa*
*Astrophysics and Cosmology Research Unit*
*University of KwaZulu-Natal, School of Mathematics*
*Durban, South Africa*
*Email: ovidiuv@yahoo.com*

*Research Associate*
*The Astronomical Institute of the Romanian Academy*
*Cutitul de Argint 5, Bucharest RO-040557 Romania*

*Research Associate*
*IMCCE – Observatoire de Paris*
*77 Avenue Denfert Rochereau, F-75014 Paris France*

MIREL BIRLAN

*IMCCE – Observatoire de Paris*
*77 Avenue Denfert Rochereau, F-75014 Paris France*
*Email: Mirel.Birlan@imcce.fr*

*Research Associate*
*The Astronomical Institute of the Romanian Academy*
*Cutitul de Argint 5, Bucharest RO-040557 Romania*

*Abstract*. The number of known Near Earth Asteroids (NEAs) and Potentially Hazardous Asteroids (PHAs) has continued to grow in the last decade. Follow-up and recovery of newly discovered objects, as well as new astrometry at second or third oppositions are necessary to improve their orbits and predict any potential collision with the Earth in the future. A project to follow-up and recovery PHAs and NEAs is proposed, using 1m class telescopes in the next two years. Two incoming runs will take place first, at Pic du Midi Observatory (France) and SAAO (South Africa), both to use 1m telescopes. Other observing runs are sought in the future. Collaborators to extend this project are welcomed.

# 1. INTRODUCTION

Follow-up observations of Near Earth Asteroids are welcomed by the astronomical community in order to recover new discovered bodies, to secure and improve their orbits and to predict their future close encounters with the Earth, including any possible collision threat well in advance.

Near Earth Asteroids (NEAs) are defined as the asteroids with a perihelion distance $q<1.3$ AU. Potentially Hazardous Asteroids (PHAs) are the NEAs having a Minimum Orbital Intersection Distance (MOID) less than 0.05 AU and absolute magnitudes $H<22$ which corresponds to objects larger than about 1 km. This limit in size represents the asteroids large enough to potentially cause a global climate disaster and threaten the continuation of human civilization (e.g., Chapman and Morrison 1994).

According to ASTORB database (Bowell 2006), there are 329,654 catalogued asteroids known today (Mar 17, 2006). According to NEO website (NASA/JPL 2005a), there are 3925 NEAs and 773 PHAs known today (Mar 17, 2006). During the last decade, these two numbers have increased dramatically.

There are about seven dedicated NEAs discovery programs in progress. Most of them have been funded and carried out in the US: LINEAR (MIT 2006, currently holding more than half of the NEAs discoveries), NEAT (NASA/JPL 2006b), Spacewatch (LPL 2006), LONEOS (Lowell Observatory 1996), and Catalina (Beshore et al 2006). Three others have been carried out elsewhere, with some interruptions due to lack of funding: Catalina South in Australia, CINEOS in Italy (Boattini et al., 2004a), and BAO in Japan (JSA 2006).

Although the annual increase of the newly discovered NEAs/PHAs has declined in the last two years, suggesting that there are fewer unknown objects detectable in the range of the present facilities, the number of known NEAs continues to grow by 200-500 new NEAs and 50-90 new PHAs discovered each year (EARN 2006).

# 2. OBSERVING BRIGHT NEAs/PHAs

Only a small fraction of the NEAs and very few PHAs are accessible to small telescopes (D<1m). About 24% of the NEAs and 22% of the PHAs have $H\leq18.0$, with 57 NEAs and 8 PHAs reaching $H\leq15$ (e.g., EARN, 2006).

Between 2002 and 2004, we carried out a follow-up program to observe a few NEAs and PHAs using the 60-cm telescope at York University Observatory in Toronto, Ontario, Canada. Due to the very light polluted sky of Toronto, the limit of this facility remained relatively modest (V≤15), which allowed minor planets to be observed only around opposition. Planning the observations in this case was simple, i.e., by searching the NEO Earth Close Approaches table maintained online by the NEO program (NASA/JPL, 2006).

Following this program, an observatory code was assigned by MPC to York University Observatory (H79) and four batches have been included in the NEODyS database and MPC Circulars (Vaduvescu, 2004b, 2004c). One paper addressed planning the observations on a small telescope (Vaduvescu, 2004a), and another presented the program at York University Observatory (Vaduvescu, 2005a).

### 3. OBSERVING FAINT NEAs/PHAs

Most of the NEAs and PHAs are faint, having absolute magnitudes H≥18.0, corresponding to sizes smaller than ~1 km (EARN, 2006). Using a 1m telescope, a good fraction of these objects are expected to be within the capabilities of a 1m class telescope (about V≤19) and can be followed at any time of the year.

In order to assert the number of objects with V≤19, we performed a database search using the MPC list "Dates of Last Observation of Unusual Minor Planets" (MPC, 2006a) to find all NEAs/PHAs with V≤19 visible between Apr 1-7, 2006. We found 30 NEAs and 7 PHAs meeting this condition, with 15 NEAs and 4 PHAs listed as "urgent", "very desirable" or "desirable" for observations. Similar searches conducted in the last year confirmed similar results. Assuming a NEA/PHA population distributed homogenously around the ecliptic, we can conclude that at least about 30 NEAs and 5 PHAs are expected to be observable weekly, using a 1m telescope.

### 4. INCOMING OBSERVING RUNS

Using 1m class telescopes, we propose a few observing runs during the next two years, depending on the available telescope and observer time.

The first observing run will take place between 15 and 29 May, 2006 at Pic du Midi Observatory, France.

A second application was submitted in March 2006 to South African Astronomical Observatory (SAAO). We hope to collect more data using the 1m telescope at SAAO by summer 2006.

Additional observing runs on 1m class telescopes are sought in the next two years, given telescope time availability in France, South Africa or elsewhere. Also, additional observers on other 1m class telescopes are welcomed!

## 5. OBSERVING STRATEGY

The main goals of the runs will be oriented toward astrometry and photometry. The reduced data will allow the improvement of the orbital elements and magnitudes. A 1m telescope located at a dark site can observe a larger number of asteroids, thus the observers have to prioritize and optimize their observing time available.

The first aim will be to recover bright newly discovered NEAs/PHAs at their first opposition. In this sense, we plan to consult, on a nightly base, the following online lists: "The NEO Confirmation Page" (MPC 2006c), "Bright Recovery Opportunities" (MPC 2006b), and "NEAObs: NEA Observation Planning Aid" (MPC 2006d). PHAs and NEAs, flagged as "Urgent" and "Desirable" will be preferred in this order.

The second aim will be to follow-up very desirable objects based on the list of "Dates of Last Observation of Unusual Minor Planets" (MPC 2006a). VIs (very important), PHAs, and NEAs (Amors, Apollo, Athens) with One-Opposition and Multi-Opposition will be preferred in this order. The MPC servers allow user to customize the lists, based on right ascension and declination limits, elongation range, V magnitude range, and observatory code.

An alternative server which can be used for double checking will be the new "HOP – Hierarchical Observing Protocol for Asteroids" (Lowell Observatory, 2006). To produce a list including all NEAs visible from a given place at present time given a limiting magnitude, the Inner-Planet Crossers flag must be set to 9, and all Other Selection Criteria must be given also the highest priority (9). Another list including NEAs at opposition is maintained by JPL (NASA/JPL, 2006).

Based on the effective observing time and number of observable objects, the third aim will be oriented toward photometry, i.e., light-curves from which to derive rotational periods of NEAs/PHAs. A list with all known asteroids rotation periods is "Minor Planet Lightcurve Parameters" (MPC, 2006e).

## 6. EPHEMERIDES AND FINDING CHARTS

We plan to use Celestial Maps 10 (Vaduvescu, Birlan and Curelaru, 2005) to produce the finding charts for the asteroids selected. Besides a few compressed catalogs available on the CD (SAO, PPM, GSC 1, Tycho-2), Celestial Maps 10 produces maps by querying online the largest stellar catalogs available today (GSC 2, USNO-A2, USNO-B1, 2MASS), via VizieR. The software includes an embedded integrator for the asteroids ephemeredes which ensures arc-second accuracy for most minor planets, using up-to-date orbital elements derived from the ASTORB database (Bowell, 2006).

Alternatively to the embedded ephemerides, one can check major online ephemerides servers, such as HORIZONS (JPL 2006) or IMCCE (IMCCE, 2006). Both take into account planetary perturbations based on modern planetary theories (DE405/406 of JPL or VSOP87 of BdL), generating accurate NEAs ephemerides to less than 1″ for most orbits.

Some NEAs/PHAs will be observed at their first or second oppositions, having large uncertainties in their orbits and consequently their ephemerides. In this case, objects having large 1-sigma uncertainty values are preferred first, although these targets are more difficult to find around their predicted positions.

## 7. EXPOSURE TIMES

Most of the NEAs/PHAs are relatively faint (V≥18), requiring longer exposure times to be able to reach a good signal to noise necessary for accurate astrometry. Most of the brightest NEAs/PHAs are expected to be observed close to opposition, thus their proper motion will be high.

The exposure times will be constraint by two variables: the apparent magnitude (which requires longer exposure times for faint targets), and the proper motion (which requires shorter exposures times in order to "freeze" the asteroid's apparent motion). Simply, for one image, the maximum exposure time in seconds is given by the ratio between the seeing, expressed in arc-seconds, to the proper motion of the asteroid, expressed in arc-seconds per second.

## 8. DATA REDUCTION

We plan to use IRAF or MIDAS to reduce astrometry and relative photometry for the observed NEAs/PHAs. Tycho-2, 2MASS, and USNO-A2 catalogs will be preferred in this order for astrometry. Collaborators might include Alin Nedelcu (The Astronomical Institute of the Romanian Academy) and Bruno Sonka (The "Admiral Vasile Urseanu Observatory" in Bucharest). Data will be reduced and sent to Minor Planet Centre within a few weeks after completion of the runs.

## 9. CONCLUSIONS

The number of known NEAs and PHAs has continued to grow in the last decade. Follow-up and recovery of newly discovered objects, as well as new data at second or third oppositions are necessary to improve their orbits and to predict any potential collision with the Earth in the future.

Some experience in NEAs/PHAs observations has been gathered by the first author in the last few years using a 60-cm telescope at York University Observatory in Toronto, Canada. More observations are proposed using 1m class telescopes in the next two years. Two incoming runs will be at Pic du Midi Observatory (France) and SAAO (South Africa), both to use 1m telescopes. Other observing runs in the future will be sought, depending on the available 1m class facilities and observer times. Collaborators to extend this project are welcomed.